\documentclass[twocolumn,8pt]{extarticle}
\usepackage{extsizes}
\usepackage{mathrsfs}
\usepackage[colorlinks=true,linkcolor=blue,urlcolor=blue,citecolor=blue]{hyperref}
\usepackage{amsthm,amssymb,amsmath,epsfig,graphics,color,verbatim,booktabs,multirow,rotating} 
\usepackage{sidecap}
\usepackage[normalem]{ulem} 
\usepackage{url} 
\usepackage{amsmath}
\usepackage{amsfonts}
\usepackage{dsfont}
\usepackage{relsize}
\usepackage{wrapfig}
\usepackage{pdfpages}
\usepackage{subfigure}
\usepackage[usenames,dvipsnames]{xcolor}
\usepackage[margin=.58in]{geometry}
\usepackage[hyperref=true,doi=false,url=false,isbn=false,backend=bibtex,style=nature]{biblatex}
\usepackage{lipsum} 
\usepackage{enumitem} 
\usepackage{verbatim} 

\renewcommand\footnotemark{$^{1}$}

\definecolor{purple}{RGB}{128,0,128}
\definecolor{green}{RGB}{0,207,0}
\definecolor{gray}{RGB}{192,192,192}

\AtEveryBibitem{
	\clearfield{month}
	\clearfield{issue}
	}

\newbibmacro{string+doiurlisbn}[1]{%
  \iffieldundef{doi}{%
    \iffieldundef{url}{%
      \iffieldundef{isbn}{%
        \iffieldundef{issn}{%
          #1%
        }{%
          \href{http://books.google.com/books?vid=ISSN\thefield{issn}}{#1}%
        }%
      }{%
        \href{http://books.google.com/books?vid=ISBN\thefield{isbn}}{#1}%
      }%
    }{%
      \href{\thefield{url}}{#1}%
    }%
  }{%
    \href{http://dx.doi.org/\thefield{doi}}{#1}%
  }%
}
\title{Magneto-optical trapping of a diatomic molecule}

\author{J.F. Barry$^{1}$, D.J. McCarron$^{1}$, E.B. Norrgard$^{1}$, M.H. Steinecker$^{1}$ \& D. DeMille\thanks{\scriptsize{\hspace{-5 mm}$^{1}$Department of Physics, Yale University, PO Box $208120$, New Haven, CT $06520$, USA.}}}

\begin{document}
\maketitle
\noindent
\textbf{Laser cooling and trapping are central to modern atomic physics. The workhorse technique in cold-atom physics is the magneto-optical trap (MOT), which combines laser cooling with a restoring force from radiation pressure. For a variety of atomic species, MOTs can capture and cool large numbers of particles to ultracold temperatures ($\boldsymbol{<\!1}$~mK); this has enabled the study of a wide range of phenomena from optical clocks to ultracold collisions whilst also serving as the ubiquitous starting point for further cooling into the regime of quantum degeneracy. Magneto-optical trapping of molecules could provide a similarly powerful starting point for the study and manipulation of ultracold molecular gases. Here, we demonstrate three-dimensional magneto-optical trapping of a diatomic molecule, strontium monofluoride (SrF), at a temperature of approximately $2.5$~mK. This method is expected to be viable for a significant number of diatomic species. Such chemical diversity is desired for the wide array of existing and proposed experiments which employ molecules for applications ranging from precision measurement, to quantum simulation and quantum information, to ultracold chemistry~\cite{Carr2009,DiRosa2004,Stuhl2008,Shuman2009,Shuman2010,Barry2012,Hummon2013,Zhelyazkova2013,Baron2014,DeMille2002,Krems2008,Micheli2006,Lane2012}.}

In a standard laser cooling experiment, the species of interest is illuminated by counter-propagating pairs of laser beams. The laser frequency is tuned just below resonance with an electronic transition, i.e. ``red-detuned'' from the transition. Due to the Doppler effect, a particle is more likely to absorb a photon from the beam that opposes the particle velocity, slowing the particle. If the subsequent (randomly directed) spontaneous emission returns the particle to the same initial state (or another state also excited by the lasers), then this process of absorption and spontaneous emission can be repeated many times, and the transition is termed a ``cycling transition.'' A damping force, described as an ``optical molasses,'' is thus provided by the laser beams. However, this cooling force does not act to spatially confine the particles. In a magneto-optical trap, cooling and confinement are produced simultaneously. To achieve this, three orthogonal pairs of laser beams are spatially overlapped with a linear magnetic field gradient in each direction (with $B=0$ at the center, a quadrupole field). For a pair of ground state and excited state Zeeman sublevels, a deviation from the trap center may induce a Zeeman shift that moves the transition closer to or further from resonance with the lasers. For a small deviation from the trap center along a given laser axis, the transition shifted closest to resonance can be driven by a particular laser polarization; this polarization is chosen for the laser counter-propagating to the direction of the deviation, while the co-propagating laser has the orthogonal polarization. Hence, on average there is a confining force restoring particles towards the center of the trap.

In this work we demonstrate magneto-optical trapping of the diatomic molecule SrF, using techniques very similar to those used for standard atomic MOTs. Anti-Helmholtz coils create a static quadrupole magnetic field ($dB_z/dz = 2\;dB_\rho/d\rho$), and pairs of circularly polarized laser beams pass through the center of this field along three orthogonal axes. The trap is loaded with pulses of SrF from a cryogenic buffer gas beam (CBGB) source \cite{Barry2011} that have been slowed using radiation pressure \cite{Barry2012}.  As we describe below, the level structure of SrF dictates that our MOT is similar to a rarely-used and poorly understood configuration of atomic MOT and exhibits some of the same atypical behavior as these atomic MOTs, such as weak confinement and slightly elevated temperature. Based on our measurements of the SrF MOT properties, we suggest means to improve on the number and density of trapped molecules reported here. Nevertheless, this work shows that standard magneto-optical trapping methods can be applied to certain molecules.

The most common atomic MOT, designated type-I, employs an $F\!\rightarrow\!F'\!=\!F\!+\!1$ cycling transition, where $F$ is the total angular momentum quantum number and the prime indicates the excited state. For a given polarization, particles in all ground-state Zeeman sublevels are optically coupled to the excited state (all states are ``bright"). The less common type-II MOT operates on an $F\!\rightarrow\!F'\!=\!F$ or \mbox{$F\!\rightarrow\! F'\!=\!F\!-\!1$} transition, where certain ``dark'' ground states sublevels are not optically coupled to the excited state. The presence of dark states reduces the spontaneous photon scattering rate; this rate can go to zero in the absence of a mechanism to ``remix" dark states with the bright states. Moreover, scattering alone does not assure a confining force; the scattering rate from the laser counter-propagating to a particle's deviation from the trap center must exceed that rate from the laser co-propagating. In type-II (but not type-I) systems, the level structure makes it possible for particles to be pumped into a state dark to the counter-propagating laser but bright to the co-propagating laser, so a confining force is not guaranteed. Consequently the damping rate and restoring force may be significantly smaller for type-II MOTs than for type-I. There appears to be no widely-accepted understanding of the mechanisms responsible for generating a restoring force in type-II MOTs. Nevertheless, type-II MOTs have been demonstrated in several atomic systems \cite{Prentiss1988,Flemming1997,Milori1997,Tiwari2008}. Due to their rotational structure, it is a generic feature of diatomic molecules that their cycling transitions correspond to a type-II MOT system.

\begin{figure} [t]
\centering
\includegraphics[width=8.9cm]{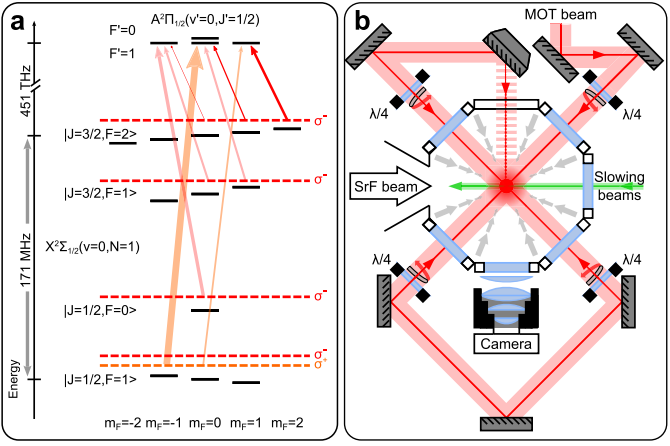}
        \caption{\textbf{a}, Optical addressing scheme for the SrF MOT. Relevant energy levels are shown for a positive $B$-field.  The $\mathcal{L}_{00}$ laser (\textcolor{Red}{\protect\rule[2pt]{.1cm}{2pt}}\;\textcolor{Red}{\protect\rule[2pt]{.1cm}{2pt}}\;\textcolor{Red}{\protect\rule[2pt]{.1cm}{2pt}}) primarily addresses the $|J\!\!=\!\!3/2,F\!\!=\!\!2\rangle$, $|J\!\!=\!\!3/2,F\!\!=\!\!1\rangle$, and $|J\!\!=\!\!1/2,F\!\!=\!\!0\rangle$ states, while the $\mathcal{L}_{00}^{\dagger}$ laser (\textcolor{Orange}{\protect\rule[2pt]{.1cm}{2pt}}\;\textcolor{Orange}{\protect\rule[2pt]{.1cm}{2pt}}\;\textcolor{Orange}{\protect\rule[2pt]{.1cm}{2pt}}) addresses the $|J\!\!=\!\!1/2,F\!\!=\!\!1\rangle$ state. Arrows show $\sigma^{-}$  transitions (\textcolor{Red}{\protect\rule[0pt]{.4cm}{4pt}}) driven by the $\mathcal{L}_{00}$ laser light and $\sigma^{+}$ transitions (\textcolor{Orange}{\protect\rule[0pt]{.4cm}{4pt}}) driven by the $\mathcal{L}^{\dagger}_{00}$ laser light. Transitions from $|J\!\!=\!\!3/2,F\!\!=\!\!1\rangle$, $|J\!\!=\!\!1/2,F\!\!=\!\!0\rangle$, and $|J\!\!=\!\!1/2,F\!\!=\!\!1\rangle$ are marked with transparent arrows for clarity; each line width is proportional to the transition strength. The lasers are drawn at the ground state energy with which they would be resonant; hence, all laser frequencies are red-detuned to the primary $J,\!F$ sublevel they address. This set of detunings is only employed for the primary trapping lasers; all repump lasers are tuned to the field-free resonance. \textbf{b}, SrF MOT experiment schematic showing the MOT (\textcolor{Red}{\protect\rule[1pt]{.4cm}{2pt}}) and slowing (\textcolor{green}{\protect\rule[1pt]{.4cm}{2pt}}) laser beam paths. The line widths indicate beam diameters, and the grey arrows illustrate the default magnetic field gradient. The $\lambda/4$ waveplates and mirrors used to create the vertical MOT beam (dashed line) are not shown.}
        \label{fig:firstfig17}
\end{figure}

We use a previously demonstrated scheme for creating a cycling transition in SrF \cite{Shuman2009,Shuman2010,Barry2012} on the $X ^2\Sigma_{1/2}^+\!\rightarrow\!A^2\Pi_{1/2}$ electronic transition (see Supplementary Information). Calculated vibrational branching fractions $b_{v'v}$, for decay of the excited state with vibrational quantum number $v'$ to the ground state with vibrational quantum number $v$, suggest that only three vibrationally-excited levels in the $X$ state ($v=1,2,3$) are significantly populated after $\sim\!10^6$ photon scattering events, corresponding to $\sim\! 1$ s of optical cycling for typical scattering rates (see below). Hence, three vibrational repumping wavelengths are expected to be sufficient to trap SrF for the $\sim\! 1$ s time scale typical of atomic MOTs. In practice we use three repump lasers, denoted $\mathcal{L}_{10}$, $\mathcal{L}_{21}$, and $\mathcal{L}_{32}$, at wavelengths $\lambda_{10}=686.0$~nm, $\lambda_{21}=685.4$~nm, and $\lambda_{32}=684.9$~nm respectively, in addition to the primary and secondary trapping lasers at $\lambda_{00}=\lambda_{00}^{\dagger}=663.3$~nm, denoted $\mathcal{L}_{00}$ and $\mathcal{L}_{00}^\dagger$ respectively. (The need for the second trapping laser is explained below.) Here $\mathcal{L}_{ij}$ denotes the laser addressing the X$(v\!=\!i)\!\rightarrow$A$(v'\!=\!j)$ transition. Radio-frequency sidebands on the $\mathcal{L}_{00}$, $\mathcal{L}_{10}$, $\mathcal{L}_{21}$, and $\mathcal{L}_{32}$ lasers address spin-rotation/hyperfine (SR/HF) structure in the X$^2\Sigma_{1/2}$ state (Fig. \ref{fig:firstfig17}a). Rotational branching is eliminated by driving an $N=1(J=3/2,1/2)\rightarrow J^\prime=1/2$ transition  \cite{Stuhl2008} where $N$ is the total angular momentum excluding electronic and nuclear spin and $J$ is the total angular momentum excluding nuclear spin. Driving these transitions optically pumps population into dark ground-state Zeeman sublevels not excited by the laser \cite{Berkeland2002}. These dark states must be remixed with the bright states for cycling to continue. In this work, remixing occurs both due to Larmor precession in the quadrupole magnetic field and due to optical pumping as molecules move through the complicated optical polarization gradients arising from the orthogonal pairs of circularly-polarized laser beams \cite{Fernandes2012}.

The optimal polarization of the trapping light depends both on the sign of the difference in magnetic moment between the ground and excited states of the cycling transition and on the orientation of the quadrupole magnetic field. In the SrF X$^2\Sigma_{1/2}(N=1)$ states, two of the four SR/HF manifolds have positive magnetic $g$-factors, one has $g=0$, and the remaining manifold has $g<0$ (Fig. \ref{fig:firstfig17}a); the A$^2\Pi_{1/2}$ state has $g \approx 0$.  The presence of both negative and positive g-factors means that laser frequencies addressing the different SR/HF manifolds must have different polarizations for optimal trapping.

The necessary trapping laser frequencies and polarizations are generated as follows. To address each SR/HF manifold, the primary $\mathcal{L}_{00}$ laser light is phase-modulated with an electro-optic modulator (EOM) (modulation frequency $f_\text{mod}=40.4$~MHz and depth $M_\text{mod}=2.6$) and combined on a polarizing beam splitter with single frequency light of the opposite polarization from the $\mathcal{L}_{00}^\dagger$ secondary trapping laser. This creates the MOT trapping light with the required frequencies and polarizations (Fig. \ref{fig:firstfig17}a). The $\mathcal{L}_{00}^\dagger$ laser is tuned closer to resonance with the $|N=1, J=1/2, F=1 \rangle$ state than the closest $\mathcal{L}_{00}$ laser sideband. With sufficient $\mathcal{L}_{00}^\dagger$ laser intensity, this ensures that molecules in the $|N\!\!=\!\!1,J\!\!=\!\!1/2,F\!\!=\!\!1\rangle$ state feel a restoring force from the $\mathcal{L}_{00}^\dagger$ light that is greater than the anti-restoring force from the nearby $\mathcal{L}_{00}$ frequency component. This additional frequency and polarization component of the light is not used for the vibrational repumping transitions since the radiative forces derived from these lasers are small. (Only $1-b_{00} \sim \! 1/50$ of the photons scattered are from the repump lasers.)

Trapping and repump light is delivered to the experiment via a single-mode polarization-maintaining fiber. Upon exiting the fiber, the $\mathcal{L}_{00}, \mathcal{L}_{10}$, and $\mathcal{L}_{21}$ lasers are vertically polarized, while the $\mathcal{L}_{00}^\dagger$ and $\mathcal{L}_{32}$ lasers are horizontally polarized. The beam of all combined frequencies and polarizations, which we refer to as the MOT light, is expanded to a $1/e^2$ intensity diameter of 14~mm. A $\lambda/2$ waveplate can rotate all polarizations by $90^\circ$. For each laser $\mathcal{L}_{ij}$, zero detuning ($\Delta_{ij}=0$) is defined as the frequency which produces maximal laser-induced fluorescence (LIF) from the molecular beam when the light is applied (and retro-reflected) perpendicular to the molecular beam.

The MOT light passes six times through the vacuum chamber (Fig. \ref{fig:firstfig17}b) and is applied at all times. Prior to the first pass, the beam is circularly polarized by a $\lambda/4$ waveplate and apertured to a $d_\lambda=23$~mm diameter. Upon exiting the chamber after this first pass, the MOT light polarization is returned to linear by a second $\lambda/4$ waveplate. This process is then repeated for the remaining radial, and thereafter, axial dimension. After the axial pass through the chamber, the MOT light is reflected back along its initial path. In this way we provide confinement in three dimensions while using the limited laser power efficiently.

The pulse of molecules from the CBGB begins with laser ablation of an SrF$_2$ target at $t=0$ ms. The slowing is applied from $t=0$ ms to $t=40$ ms (see Supplementary Information). Molecules in the trapping region are detected via LIF from the X$\rightarrow$A cycling transition at $\lambda_{00}=663.3$~nm and imaged onto a CCD (see Supplementary Information). The camera field-of-view encompasses the majority of the trapping region and is approximately centered on the zero of the quadrupole field. Unless otherwise noted, MOT imaging starts at time ${t_\text{im}=60}$~ms after ablation, the camera exposure duration is $t_\text{exp}=60$~ms, and the signal is integrated  over the entire camera field-of-view. The imaging start time and duration are chosen so that the vast majority of LIF recorded ($\sim$90\%) comes from  trapped molecules rather than the temporal tail of the slowed molecular beam pulse.

Realization of magneto-optical trapping results in increased LIF from a small area in the trapping region near the B-field zero. This localized LIF persists for an increased duration compared to the spatially broad LIF from the untrapped, slowed molecular beam and suggests that molecules are confined in this region. To observe the MOT, the $\mathcal{L}_{00}, \mathcal{L}_{00}^\dagger, \mathcal{L}_{10}$, and $\mathcal{L}_{21}$ lasers must be present with the proper detunings (and, for the trapping lasers, polarizations), the B-field gradient must be present $(dB_{z}/dz \neq 0)$, and the laser slowing must be applied. The $\mathcal{L}_{32}$ laser is not necessary to observe the MOT but results in increased LIF. Maximum LIF is observed when $\Delta_{00} = \Delta_{00}^\dagger = - 1.2\Gamma\approx2\pi\times 8$~MHz and $dB_{z}/dz = 15$~G/cm (see Supplementary Information); these parameters are similar to those for standard atomic MOTs. Unless stated otherwise, measurements are made with these default parameters.

\begin{figure}
\centering
\includegraphics[width=8.9cm]{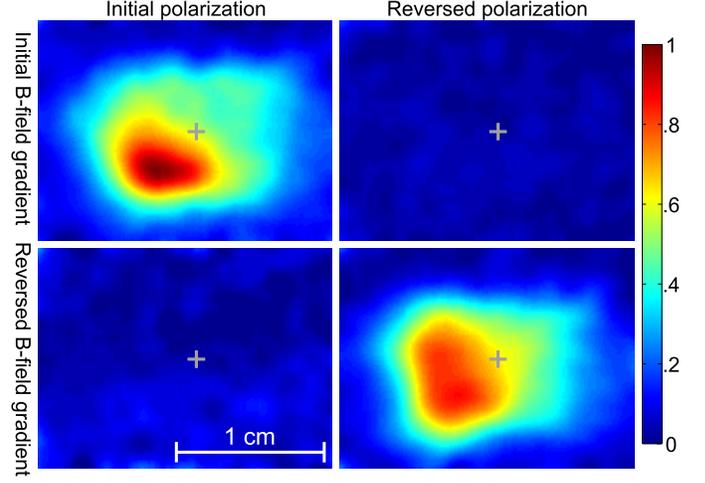}
\caption{Magneto-optical trapping of SrF. Shown are images of LIF in the trapping region for different polarizations and signs of $dB_{z}/dz$ as described in the main text. Each image is an average over 600 pulses. The grey crosses mark the position of the magnetic field zero on each image. The intensity scale is in arbitrary units.}  \label{fig:quadrantfigure}
\end{figure}

The proper polarization for the trapping light depends on the sign of $dB_{z}/dz$. Reversing either the sign of $dB_{z}/dz$ or the circular polarization of the MOT trapping light should create an anti-restoring force and prevent MOT formation. Reversing both the sign of $dB_{z}/dz$ and the polarization in tandem, however, should return the system to a restoring configuration, and the MOT should be realized again. We observe the expected behavior for these four states of the system as shown in Fig. \ref{fig:quadrantfigure}, confirming magneto-optical trapping of SrF molecules. From these images we also determine the MOT cloud position and size by fitting the LIF intensity profile to a 2D Gaussian with adjustable center, width, and rotation angle; we find typical Gaussian (r.m.s.) widths of $\rho_{\rm{rms}} = 4.1(1)$~mm (radial) and $z_{\rm{rms}}=2.6(1)$~mm (axial).

\begin{figure*}[t]
\centering
\includegraphics[width=18.3cm]{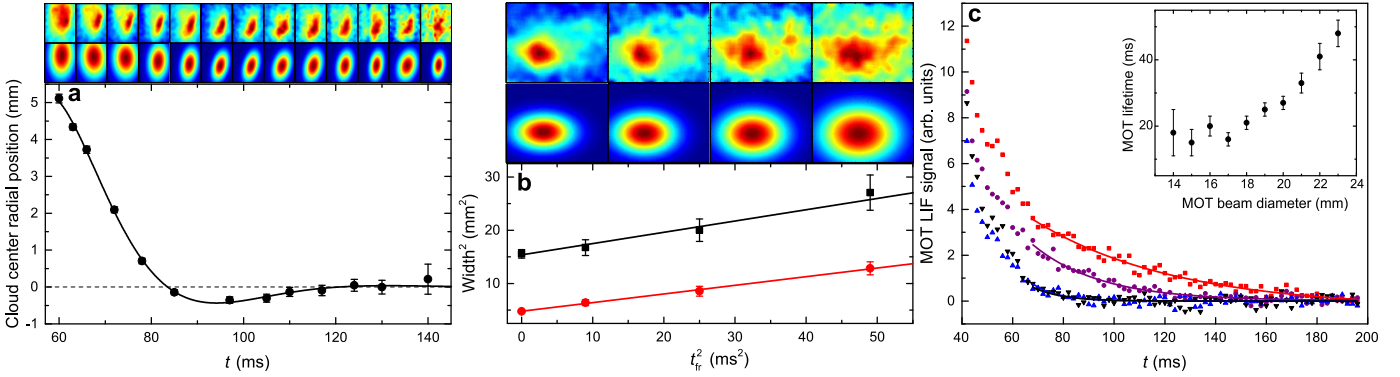}
\caption{\textbf{a}, MOT cloud response to rapid displacement of the trap center. Shown are MOT LIF images, each averaged over 1600 pulses (top), 2D Gaussian fits (middle), and the extracted radial position (bottom) as a function of time. (In these images, the radial axis is oriented vertically.) The fit is to the motion of a damped harmonic oscillator (see main text). Zero is set at the position of the MOT with no displacement. \textbf{b}, Free expansion of the MOT following release. Shown are MOT LIF images, each averaged over 2000 pulses (top), 2D Gaussian fits (middle), and measured MOT radial (\footnotesize{\color{black}$\blacksquare$}\normalsize) and axial ({\color{red}$\bullet$}) widths vs. free expansion time (bottom); see main text for fit function. In both \textbf{a} and \textbf{b}, images are rescaled to the maximum value at each time. \textbf{c}, LIF in the trapping region vs. time $t$ for MOT  with (\tiny{\color{red}$\blacksquare$}\normalsize) and without (\footnotesize{\color{purple}$\bullet$}\normalsize) the $\mathcal{L}_{32}$ repump laser, for untrapped molasses (\small{\color{black}$\blacktriangledown$}\normalsize), and for damping/anti-restoring (\small{\color{blue}$\blacktriangle$}\normalsize) configurations (see Supplementary Information). Overlaid are single exponential fits. Inset: MOT lifetime vs. MOT laser beam diameter. All error bars show the $\pm 1\sigma$ confidence interval.}  \label{fig:allcombineddata}
\end{figure*}

To probe the confining and damping forces in the MOT, the molecular cloud's response to a rapid displacement of the trap position is measured. A magnetic bias field applied from $t=0$~ms offsets the MOT center by $\sim \! 5$~mm radially. The bias field is switched off at $t_\text{off}=58$~ms, releasing the trapped molecules into the unbiased potential. The molecular cloud's position is measured as a function of time (Fig. \ref{fig:allcombineddata}a); see Supplementary Information. The cloud exhibits damped harmonic motion as it moves towards the equilibrium position, with oscillation frequency $\omega_{\rho}= 2\pi\times17.2(6)$~Hz and damping coefficient $\alpha=2.4(2)\times10^{-23}$~kg/s.  The radial spring constant is given by $\kappa_{\rho}=\omega_\rho^{2}m_{\rm{SrF}} = 2.1(1)\times10^{-21}$~N/m, where $m_{\rm{SrF}}$ is the mass. With $\kappa_\rho$ and the measured radial width ($\rho_{\rm{rms}}$), the equipartition theorem is used to find the radial MOT temperature, $T_\rho= \kappa_\rho \rho_\text{rms}^2/k_B = 2.5(2)$ mK, where $k_B$ is the Boltzmann constant. This temperature corresponds to an r.m.s. molecular speed $v_{\rm{rms}}=0.77(3)$~m/s and, given the measured value of $\alpha$, a typical damping force $F=-\alpha v_{\rm{rms}} = 1.8(1)\times10^{-23}$~N. We are unable to repeat this measurement to determine the axial spring constant ($\kappa_{z}$) as the MOT loading is impeded when the trap center is offset axially (vertically) away from the slowed molecular beam. However, for a standard atomic MOT in a quadrupole field, the relation $\kappa_{z} = 2\kappa_{\rho}$ holds \cite{Metcalf1999}, which suggests $\kappa_{z}=4.1(3)\times10^{-21}$~N/m and $\omega_{z}= 2\pi\times24.3(9)$~Hz. The measured MOT axial width $z_{\rm{rms}}$ then corresponds to an axial temperature $T_{z}=2.0(1)$~mK.

To verify the MOT temperature, ballistic expansion measurements are performed. Trapped molecules are released at $t_\text{rel}\!=\!90$~ms (by extinguishing the $\mathcal{L}_{21}$ laser). After a free expansion time $t_{\text{fr}}$, the $\mathcal{L}_{21}$ light is restored, and the resulting LIF is imaged onto the CCD. A short imaging time must be used to accurately determine the expanded cloud's size using this LIF-based method, since the trapping light used for the LIF imaging will eventually recapture and compress the cloud. The camera exposure interval over which the expanding cloud is illuminated and imaged, $t_{\text{exp}}=5$~ms, is short enough that this mechanical effect of the light is sufficiently small, i.e. $t_{\text{exp}}\ll 2 \pi / \omega_{z}$. For an initial Gaussian spatial distribution and a Boltzmann distribution of velocities (with no correlation between position and velocity), the widths $z_\text{rms}$ and $\rho_\text{rms}$ of the expanding cloud are given by
\begin{equation}
\rho_\text{rms}^{2} = \frac{k_{B}T_{\rho}}{m_{\rm{SrF}}}\Bigg(\frac{1}{\omega_\rho^{2}}+t_{\rm{fr}}^2\Bigg) \;\;\;\;\; \text{and} \;\;\;\;\;  z_\text{rms}^{2} = \frac{k_{B}T_{z}}{m_{\rm{SrF}}}\Bigg(\frac{1}{\omega_z^{2}}+t_{\rm{fr}}^2\Bigg).
\end{equation}
The data and associated fits are shown in Fig. \ref{fig:allcombineddata}b. The slopes of the fits give the temperatures, which are then used with the intercepts to determine $\omega_{\rho}$ and $\omega_{z}$. This treatment of the data plots the measured cloud width (which is an average over the camera exposure duration, $t=t_\text{im}\rightarrow t_\text{im}+t_\text{exp}$) at the start time ($t=t_\text{im}=t_\text{rel}+t_\text{fr}$) of each illumination/imaging period. Monte Carlo simulations for the measured trap frequencies suggest that the cloud continues to expand during the short illumination interval, and therefore the extracted width is an upper bound for the actual width at $t=t_\text{im}$. Hence this treatment of the data yields upper limits on the MOT temperature. The fits give $T_{\rho}\le2.7(3)$~mK and $T_{z}\le2.1(1)$~mK, with trap frequencies $\omega_{\rho}=2\pi\times19(1)$~Hz and $\omega_{z}=2\pi\times29(1)$~Hz, corresponding to spring constants $\kappa_{\rho}=2.5(3)\times10^{-21}$~N/m and $\kappa_{z}=5.9(4)\times10^{-21}$~N/m. These values are in good agreement with the values from the MOT oscillation measurement. These spring constants are two to three orders of magnitude smaller than for typical type-I atomic MOTs \cite{Wallace1994} and approximately one order of magnitude smaller than reported values for atomic type-II MOTs \cite{Tiwari2008}, though measurements of the spring constants in type-II atomic MOTs are so few that ``typical" behavior is difficult to characterize.

Another independent measurement of the MOT temperature $T_\text{MOT}$ is performed using the release-and-recapture method \cite{Lett1988}. Here, trapped molecules are released and then expand freely for a variable time $t_\text{fr}$ before the trap is turned back on, recapturing a fraction of the initial molecules. It is assumed that molecules expand isotropically and that molecules that travel outside a capture radius $r_\text{cap}$ are lost. An isotropic temperature $T_\text{iso}$ is inferred from the fraction of molecules recaptured vs. $t_\text{fr}$ (see Supplementary Information). For this model we assume $r_\text{cap}= d_\lambda/2$, the radius of the MOT beam; this assumption sets the extracted value of $T_\text{iso}$ as an upper limit. This method yields the isotropic temperature $T_\text{iso} < 2.7(6)$~mK, in good agreement with the geometric mean of the axial and radial MOT temperatures derived from the MOT oscillation measurement $(T_{z}T_{\rho}^{2})^{1/3}=T_\text{MOT}=2.3(4)$~mK; this is roughly an order of magnitude greater than the SrF Doppler temperature, $T_{D}=\hbar \Gamma/(2 k_{B})= 160~\mu$K, where $\Gamma = 2\pi \times 7$~MHz is the natural linewidth. Temperatures well above the Doppler temperature are also reported in work with atomic type-II MOTs \cite{Tiwari2008}.

Measurement of the spontaneous photon scattering rate for trapped molecules, $R_\text{sc}$, allows the number of trapped molecules $N_\text{MOT}$ to be determined via fluorescence detection (see Supplementary Information). The value of $R_\text{sc}$ is measured by blocking the $\mathcal{L}_{21}$ repump light at $t_\text{bl}=58.6$~ms and observing the LIF decay constant, denoted $\tau_\text{v=2}$, as molecules are optically pumped into the now-dark ${\rm{X}^2\Sigma_{1/2}(v=2)}$ state. From the observed value $\tau_\text{v=2}=0.66_{-0.32}^{+0.70}$ ms and the calculated vibrational branching ratio $b_{02}\approx 0.0004$, we estimate $R_\text{sc} = 1/(b_{02}\tau_\text{v=2})\approx 4.3_{-2.2}^{+4.1}\times10^6$~s$^{-1}$. Based on the efficiency of the LIF detection system, measured to be $\sim\!0.8\%$, the MOT population is estimated at $N_\text{MOT}\!\approx\!400$ SrF molecules, corresponding to a peak trap density of $\approx \! 800$~cm$^{-3}$.

The measured value of $R_\text{sc}$ makes it possible to estimate a lower limit on the MOT heating rate (additional heating may be caused by stimulated emission, which is neglected here). This minimum heating rate is $(dE/dt)_\text{h} = 2R_\text{sc} E_\text{r}$ where $E$ is the molecule energy and $E_\text{r}= (\hbar k)^{2}/(2m_{\rm{SrF}})$ is the photon recoil energy ($k=2\pi/\lambda_{00}$ is the wavenumber). By equating the heating rate to the rate of cooling from velocity damping, $(dE/dt)_\text{c} = -2\alpha k_{B}T_\text{MOT}/m_\text{SrF}$ \cite{Metcalf1999},  an independent lower limit on the damping coefficient $\alpha$ is obtained. We find $\alpha\ge5^{+5}_{-3}\times10^{-23}$~kg/s, in agreement with the value from the MOT oscillation measurement.

The MOT lifetime, $\tau_{\rm{MOT}}$, is obtained by measuring LIF in the trapping region as a function of time and fitting a single exponential decay curve to the data after $t=67$~ms, as shown in Fig. \ref{fig:allcombineddata}c. This start time avoids significant contributions to the LIF signal from the slowed but ultimately untrapped part of the molecular beam. We find $\tau_\text{MOT} =56(4)$~ms, significantly shorter than is typically seen in atomic MOTs. When the $\mathcal{L}_{32}$ repump laser is not present $\tau_{\rm{MOT}}=27(2)$~ms. The values of $\tau_\text{MOT}$ with and without the $\mathcal{L}_{32}$ repump laser allow the loss rate into the X$^2\Sigma_{1/2}(v=3)$ state to be isolated. This, together with the vibrational branching ratio $b_{03}$ into $v=3$, yields an independent measurement of the scattering rate $R_\text{sc} \approx 2 \times10^{6}$~s$^{-1}$, with uncertainty of $\sim\!100$\% due to the large uncertainty in $b_{03}$ (see Supplementary Information).

It is verified that neither collisions with ballistic helium from the buffer-gas beam nor collisions with background gases are the primary loss mechanism from the trap. Optical pumping into the dark X$^2\Sigma_{1/2}(v=4)$ state would result in $\tau_{\rm{MOT}}\approx 1$ s for the measured value of $R_\text{sc}$, and off-resonant excitation populating dark rotational levels is found to be insignificant (see Supplementary Information).

The strikingly low restoring force measured suggests another possible explanation for the low value of $\tau_\text{MOT}$: the trap depth is not large compared to $k_{B}T_\text{MOT}$, as in typical atomic MOTs, so a significant fraction of molecules can escape the trap simply by being in the high-energy tail of the Boltzmann distribution. The MOT trap depth $U_\text{MOT}$ can be estimated using $U_\text{MOT}=\frac{1}{2}\kappa_{\rho} (d_\lambda/2)^{2}$, assuming that $\kappa_{\rho}$ is constant to the edges of the MOT beam. This gives $U_\text{MOT}/k_{B}=10(1)$~mK $\approx\!4 T_{\rm{MOT}}$, in contrast to atomic MOTs where $U_\text{MOT}/k_{B}\!\approx\! 1$~K $\!\approx\! 1000 T_\text{MOT}$. We presume that rapid molecule-light interactions maintain a constant temperature in the MOT, leading to continuous loss rather than evaporative cooling as in a conservative trap. A simple model for the rate of particle escape under these conditions lends credence to this explanation for the short MOT lifetime (see Supplementary Information). Additional support comes from the observation that $\tau_\text{MOT}$ depends strongly on the MOT beam diameter (Fig. \ref{fig:allcombineddata}c, inset). Reducing the beam diameter $d_\lambda$ from 23~mm to 21~mm (a $<\!1\%$ decrease in power) reduces $\tau_\text{MOT}$ by $\sim\!30\%$. We are unaware of any other trap loss mechanism that might exhibit this behavior.

For our cycling transition, the maximum restoring force $F_\text{max} = \kappa_{z} d_\lambda/2$ corresponds to scattering $F_\text{max}/(\hbar k) = R_\text{con}=5(2) \times 10^4$~s$^{-1}$ confining photons from a single MOT beam, only $\sim\!1\%$ of $R_\text{sc}$. The small value of $R_\text{con}/R_\text{sc}$ may be understood qualitatively by noting that in a simple 1-dimensional model, the angular momentum level structure of our system $(J=3/2,1/2 \rightarrow J^\prime = 1/2)$ ensures that each photon scattered in the ``correct'' (confining) direction on average must be followed by a photon scattered in the ``incorrect'' (anti-confining) direction in order to resume optical cycling \cite{Nasyrov2001}. In 3D, with complicated polarization gradients and other means of remixing, this relation no longer holds exactly. Nonetheless, the mechanism behind the slight bias of scattering events towards the trap center that leads to the weak, yet nonzero confining force is not well understood. This same type of level structure is also the defining characteristic of atomic type-II MOTs, which exhibit qualitatively similar characteristics to our SrF MOT (extended spatial extent, elevated temperature, etc.) although with reported stronger confinement \cite{Prentiss1988,Flemming1997,Milori1997,Tiwari2008}. Hence the weak trapping and only moderately low temperature observed in our SrF MOT are believed to be due to the angular momentum level structure rather than any other issues related specifically to using a diatomic molecule rather than an atom. Despite these limitations, our method succeeds in trapping and cooling molecules to the lowest temperature reported for any direct-cooling method to date.

Future work is expected to allow substantial increases in the density and the number of molecules trapped. For example, the trappable flux may be increased by implementing a more efficient slowing method \cite{Eyler2012} or by transversely confining the molecular beam as it is slowed \cite{DeMille2013}. A variety of methods may enable increased trap depth by increasing the fraction of scattered photons contributing to the confining force ($R_\text{con}/R_\text{sc}$), which could in turn increase trap lifetime, capture velocity, density, and number of molecules trapped. This could be accomplished, e.g., by rapid synchronous reversing of the MOT magnetic field gradient and the laser circular polarizations, as recently demonstrated in 2D magneto-optic compression of a molecular beam \cite{Hummon2013}, or alternatively by using a microwave electric field to pump molecules in anti-trapped Zeeman sublevels into trapped levels by driving transitions through the X$(v\!=\!0,N\!=\!0)$ rotational state \cite{Stuhl2008,Shuman2009}.

Although magneto-optical trapping of molecules is in its infancy, our results demonstrate that this technique can be applied in a straightforward way to a significant number of diatomic species \cite{DiRosa2004,Stuhl2008,Lane2012}. The MOT has proved indispensable for cooling and trapping many atomic species; with further development, we expect that it may prove similarly useful for producing ultracold gases of diatomic molecules. Such an advance is expected to enable a wide range of new experiments including tests of the Standard Model of particle physics \cite{Isaev2010,Hunter2012,Tarbutt2013}, sensitive searches for variations of fundamental constants \cite{Chin2009}, production of exotic ultracold atomic species \cite{Lane2012}, and studies of novel chemical dynamics in the ultracold temperature regime \cite{Stuhl2008,Lane2012}. 
\newline

\noindent \textbf{Acknowledgements} We thank E.R. Edwards for contributions towards the construction of the experiment. We acknowledge funding from AFOSR (MURI), ARO, and ARO (MURI). E.B.N. acknowledges funding from the NSF GRFP.
\\
\\
\noindent \textbf{Author Contributions} All authors contributed to the experiment, the analysis of the results and the writing of the manuscript.
\\
\printbibliography
\clearpage

\section*{SUPPLEMENTARY INFORMATION}

\section*{Cycling scheme and level structure}
\begin{figure*}
\centering
\includegraphics[width=18.3cm]
{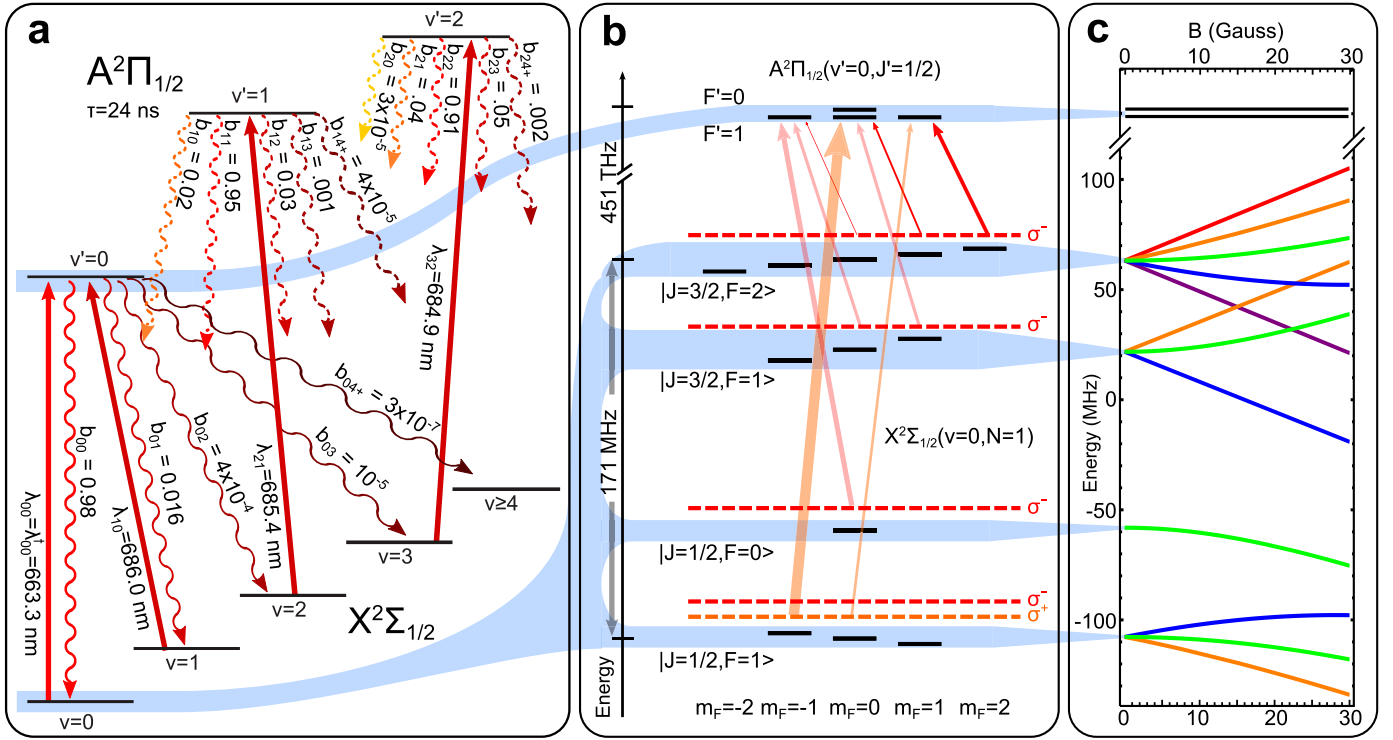}
\caption{\textbf{a}, Vibrational branching in SrF. Solid upward lines denote transitions driven by the MOT lasers. Spontaneous decays from the A$^2\Pi_{1/2}(v'=0)$ state (solid wavy) and A$^2\Pi_{1/2}(v'=1,2)$ states (dashed wavy) are governed by the vibrational branching fractions $b_{0v}$, $b_{1v}$, and $b_{2v}$, as shown. \textbf{b} Optical addressing scheme for the SrF MOT presented and discussed in the main text. \textbf{c}, Energy levels of the $X^2\Sigma_{1/2}(v=0,N=1)$ state versus $B$. Energy levels are labeled by their $m_F$ value with $m_F=2$ {(\color{red}\LARGE\textbf{-}\normalsize\color{black})}, $m_F=1$ {(\color{orange}\LARGE\textbf{-}\normalsize\color{black})}, $m_F=0$ {(\color{green}\LARGE\textbf{-}\normalsize\color{black})}, $m_F=-1$ {(\color{blue}\LARGE\textbf{-}\normalsize\color{black})} , $m_F=-2$ {(\color{Fuchsia}\LARGE\textbf{-}\normalsize\color{black})}.}
\label{fig:firstfig}
\end{figure*}

The $X ^2\Sigma_{1/2}^+\!\rightarrow\!A^2\Pi_{1/2}$ electronic transition employed in this work has a lifetime $\tau_{A}=1/\Gamma=24.1$~ns. The vibrational branching for this excited state is shown in Fig. \ref{fig:firstfig}a and dictates that only four lasers are required to cycle $\gtrsim10^{6}$ photons. Uncertainties in the calculated vibrational branching fractions $b_{v'v}$ stem largely from uncertainties in the molecular constants for the $A^2\Pi_{1/2}$ state used to calculate Franck-Condon factors. Although the errors in these constants are small, the resulting fractional uncertainty in calculated values of $b_{v'v}$ may be significant for off-diagonal terms ($v\ne v'$) where Frank-Condon factors are strongly suppressed and vibrational branching ratios are small \cite{Barry2013}.

The SR/HF structure for the $X^{2}\Sigma_{1/2}(v=0,N=1)$ state in the presence of a weak magnetic field  is shown in Fig. \ref{fig:firstfig}b. For the three repump lasers ($\mathcal{L}_{10}$, $\mathcal{L}_{21}$, and $\mathcal{L}_{32}$), the modulation frequency $f_\text{mod}=42.5$~MHz is chosen so the first- and second-order sidebands address all four SR/HF transitions. The value of $f_\text{mod}=40.4$~MHz for the $\mathcal{L}_{00}$ light is chosen to minimize the root-mean-squared (r.m.s.) value of the detuning for the upper three SR/HF levels at $B=0$~G, while a separate laser addresses the lowest SR/HF level. The need for an additional trapping laser, with the opposite polarization, is discussed in the main text. A Breit-Rabi diagram showing the energy dependence of each sublevel vs. magnetic field is shown in Fig. \ref{fig:firstfig}c. The level crossings in the range $B=15$ to $25$~G may limit the effective trap radius for a given $B$-field gradient since, at sufficiently high fields, the trap light frequency addressing the $|J\!\!=\!\!3/2,F\!\!=\!\!1\rangle$ manifold becomes anti-trapping for the $|J\!\!=\!\!3/2,F\!\!=\!\!2\rangle$ manifold. Note that other trapping/anti-trapping level crossings are located at higher magnetic bias fields.

\section*{Radiation pressure slowing}
The molecular beam is slowed by three lasers, denoted $\mathcal{L}_{00}^\text{s}$, $\mathcal{L}_{10}^\text{s}$, and $\mathcal{L}_{21}^\text{s}$ (where the ``s'' superscript indicates slowing), which have powers of 205~mW, 185~mW and 35~mW respectively. The $\mathcal{L}_{00}^\text{s}$  and $\mathcal{L}_{21}^\text{s}$ lasers are horizontally polarized while the  $\mathcal{L}_{10}^\text{s}$ laser is vertically polarized. These lasers are spatially overlapped to produce a single beam with $1/e^2$ intensity diameter d $\sim$\! 3~mm, applied counter-propagating to the molecular beam. A uniform field $B^\text{s} \! \sim \! 9$~G is applied at an angle $\theta=45^\circ$ relative to the linear polarizations of the lasers over the distance -200~mm~$\lesssim\! z'\! \lesssim$~1600~mm, where $z'=0$ marks the exit of the cell in the CBGB source and $z'$ denotes the downstream distance along the molecular beam. The magnetic field for the slowing is non-zero only when the slowing lasers are applied.

\begin{figure}
\includegraphics[width=8.9cm]{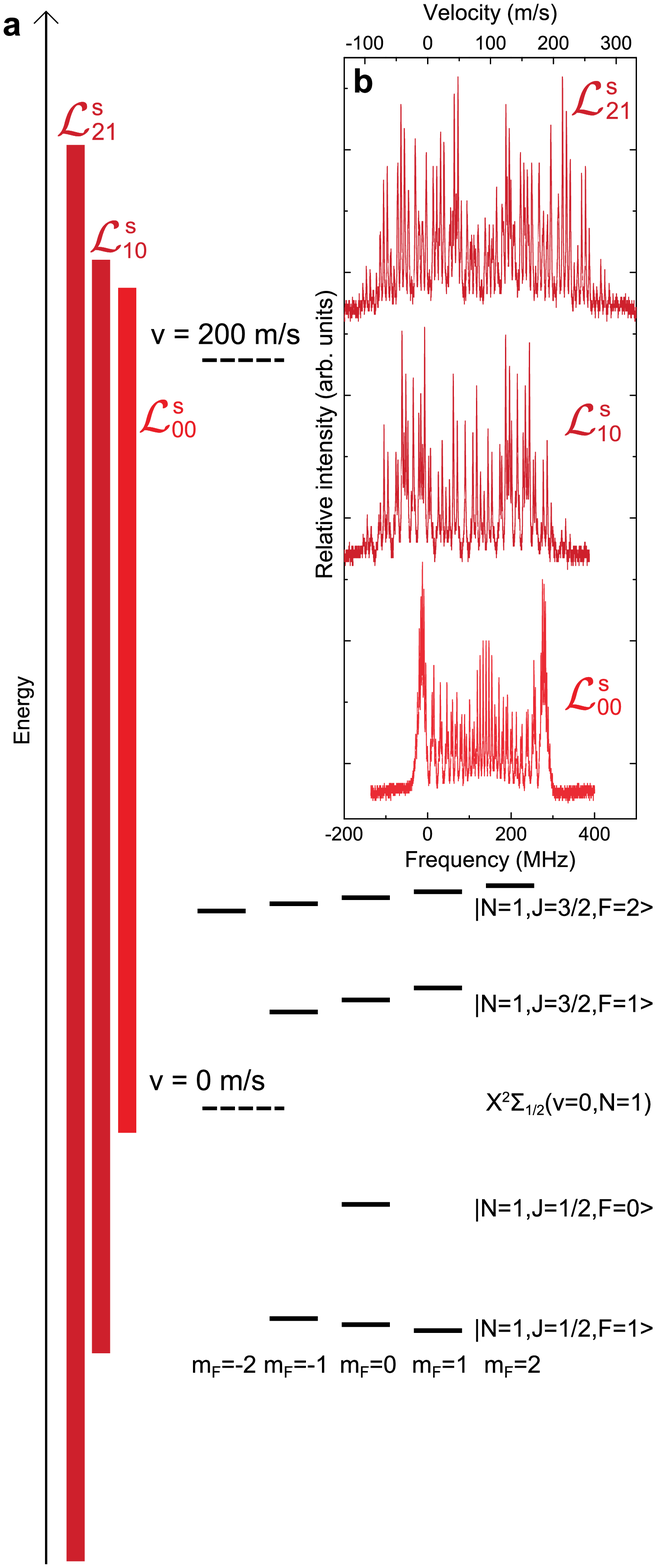}
\caption{\textbf{a}, Scale diagram showing the frequency extent of the $\mathcal{L}_{00}^s$, $\mathcal{L}_{10}^\text{s}$, and $\mathcal{L}_{21}^s$ slowing lasers relative to the four SR/HF manifolds of SrF's X$^2\Sigma_{1/2}(v=0,1,2;N=1)$ states. The relative splittings of the four SR/HF levels in the X$^2\Sigma_{1/2}(N=1)$ state are the same to within $\sim \! 1$~MHz for $v=0,1,2$ \cite{Barry2013}. The dashed lines mark the centers of the $N=1$ SR/HF levels for the labelled velocity, and the level structure shown corresponds to $v=0$~m/s. \textbf{b}, Optimized spectral profiles of the three slowing lasers. The top scale shows velocity for a Doppler shift equivalent to the frequency labelled on the bottom scale. The $\mathcal{L}_{00}^s$ light is modulated via a fiber EOM with $f_\text{mod}=3.5$~MHz. The $\mathcal{L}_{10}^\text{s}$, and $\mathcal{L}_{21}^s$ lasers are each modulated by passing through two bulk EOMs with resonant frequencies at $\approx40$~MHz and $\approx9$~MHz.}
\label{fig:v00splot}
\end{figure}

The properties of the slowing (laser center frequencies and frequency extents, application time and duration of the slowing, and value of $B^\text{s}$) are optimized by imaging the MOT after the slowed molecular beam pulse has fully subsided (here from $t=80$ to $t=110$~ms). The optimized frequency detunings of the $\mathcal{L}_{00}^\text{s}$, $\mathcal{L}_{10}^\text{s}$, and $\mathcal{L}_{21}^\text{s}$ lasers are $\Delta_{00}^{\text{s}}=-161$~MHz, $\Delta_{10}^{\text{s}}= -122$~MHz, and $\Delta_{21}^{\text{s}}=-103$~MHz. The spectra are broadened to address a wide range of Doppler shifts associated with the broad velocity spread from the CBGB source. The spectral widths are $340$~MHz, $440$~MHz, and $570$~MHz for the $\mathcal{L}_{00}^\text{s}$, $\mathcal{L}_{10}^\text{s}$, and $\mathcal{L}_{21}^\text{s}$ lasers respectively (Fig. \ref{fig:v00splot}). Further details on the slowing can be found in Refs. \cite{Barry2012,Barry2013}.

\begin{figure}
\centering
\includegraphics[width=8.9cm]{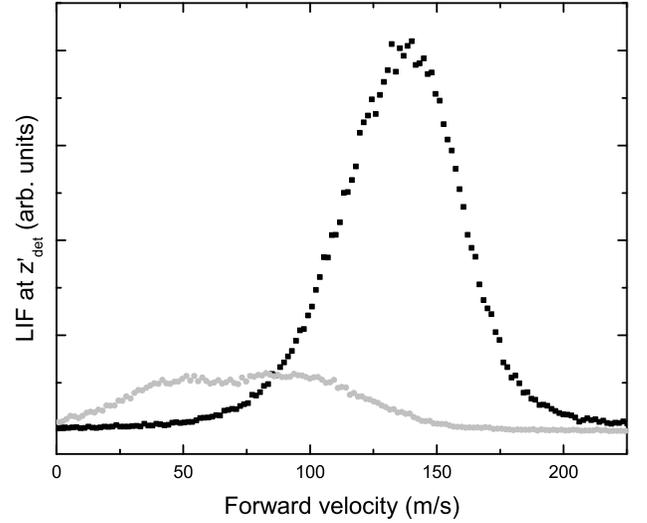}
\caption{Examples of slowed (\footnotesize{\color{gray}$\bullet$}\normalsize) and unslowed (\footnotesize{\color{black}$\blacksquare$}\normalsize) velocity profiles of the molecular beam detected upstream from the trapping region at $z'_\text{det}=1076$~mm. These profiles are for the optimized slowing conditions that produce the largest MOTs as discussed in the main text.}
\label{fig:exampleslowing}
\end{figure}

Fig. \ref{fig:exampleslowing} shows a sample slowed velocity profile used to load the MOT, along with the unslowed velocity profile of the source; both profiles are detected upstream of the trapping region at $z'_\text{det}=1076$~mm.


\section*{Trapping region}
The trapping region for the MOT is centered at $z'=1382$~mm and is separated from the beam propagation region by a differential pumping tube (127~mm long, 12.7~mm diameter) beginning at $z' \! \approx \! 900$~mm. In the trapping region, the pressure of all background gas excluding He is $P_\text{BG} \approx 4\times 10^{-10}$~Torr while the helium background pressure is $P_\text{He} \approx 2\times 10^{-9}$~Torr.

\section*{MOT optimization}
\begin{figure}
\includegraphics[width=8.9cm]{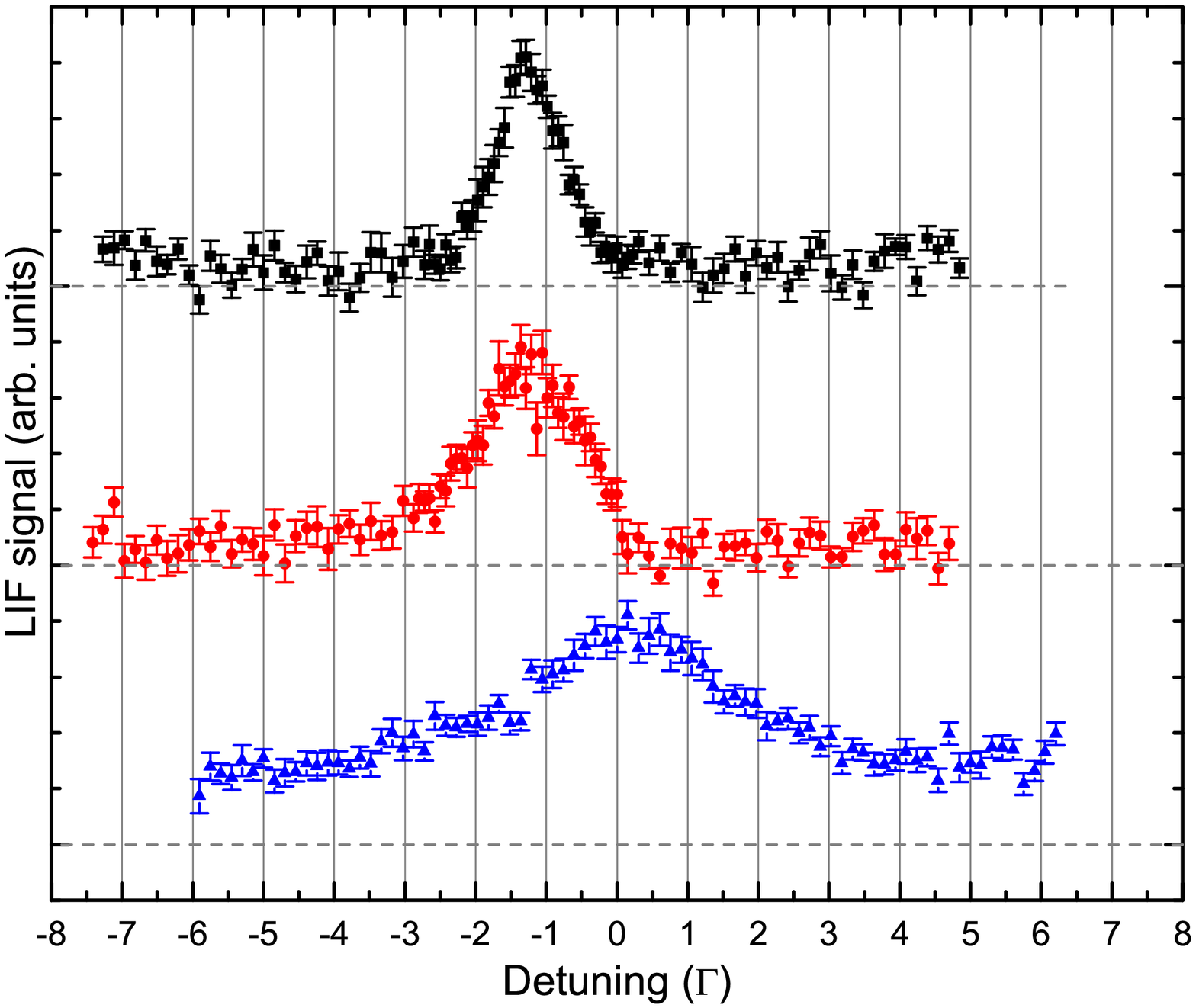}
\caption{LIF in the trapping region vs. detuning when $\Delta_{00}$ and $\Delta_{00}^\dagger$ are varied together (\footnotesize{\color{black}$\blacksquare$}\normalsize), when $\Delta_{00}^\dagger$ is varied alone (\small{\color{red}$\bullet$}\normalsize), and when $\Delta_{10}$ is varied (\small{\color{blue}$\blacktriangle$}\normalsize). As expected (and typically observed for atomic MOTs), the SrF MOT operates over a fairly narrow range of red detuning values for the trapping lasers but requires only that the repump laser be sufficiently near resonance.}
\label{fig:LIFvsdetuning}
\end{figure}

The optimum magnetic field gradient for the MOT is $dB_{z}/dz=15$~G/cm; the MOT is visible between 4 and 30~G/cm. The MOT is sensitive to the values of the laser detunings $\Delta_{00}$ and $\Delta_{00}^\dagger$ (and less sensitive to the value of $\Delta_{10}$) as shown in Fig. \ref{fig:LIFvsdetuning}. The MOT is insensitive to the detunings of the $\mathcal{L}_{21}$ and $\mathcal{L}_{32}$ lasers.

As discussed in Refs. \cite{Barry2013,Tarbutt2013}, laser cooling schemes with a large number of resolved ground states can require significantly more power than those employing a two-level system with the same wavelength and electronic excited state lifetime. Briefly, a $F\!=\!1\rightarrow F'\!=\!0$ type transition will have a saturation intensity $3\times$ higher than the saturation intensity for a two-level system of the same wavelength and lifetime. Resolved ground state energy levels also increase the required intensity by dictating that total laser power be divided up among several frequencies, each driving a weaker transition. Hence, as anticipated, our molecular MOT requires substantially more laser power than standard atomic MOTs. Upon exiting the MOT fiber, the $\mathcal{L}_{00}, \mathcal{L}_{00}^\dagger, \mathcal{L}_{10}$, $\mathcal{L}_{21}$, and $\mathcal{L}_{32}$ laser powers are typically 210~mW, 50~mW, 170~mW, 5~mW, and 3~mW respectively.

\section*{Spontaneous scattering rate for trapped molecules}
To measure the spontaneous photon scattering rate, $R_\text{sc}$, LIF is recorded as a function of imaging start time $t_\text{im}$, which is scanned from $t_\text{im}=54$~ms to $t_\text{im}=62$~ms. The $\mathcal{L}_{21}$ repump light is blocked at $t_\text{bl}=58.6$~ms (see main text). The finite duration of the camera exposure, $t_\text{exp}=1$~ms, results in a recorded LIF signal $Y(t)$ that is a convolution of the real instantaneous LIF intensity, denoted $X(t)$, and the camera exposure time, i.e.,
\begin{equation}
Y(t) = \int_{t_\text{im}}^{t_\text{im}+t_\text{exp}} X(t')dt'.
\end{equation}
Given the comparatively long unperturbed MOT lifetime, $X(t)$ is modeled as a linear function prior to the blocking of the $\mathcal{L}_{21}$ repump light ($t=54$ to $t_{0}=t_\text{bl}-t_\text{exp}=57.6$~ms), followed by (from $t_{0}$) an exponential decay plus an additional linear background term (to account for LIF from the tail of the slowed but untrapped molecular beam); this background term is deduced from a fit to the data from $t = 59$ to $t=62$~ms. This function has the form
\begin{align}
X(t) = (m_\text{MOT}t+&c_\text{MOT})H(t_{0}-t)\nonumber \\
& - (D_{0}\,e^{-(t-t_{0})/\tau_\text{v=2}} + m_\text{bg}t+c_\text{bg})H(t-t_{0}),
\end{align}
where $m_\text{MOT}$ and $c_\text{MOT}$ ($m_\text{bg}$ and $c_\text{bg}$) are the gradient and intercept respectively of the linear fit to the LIF from the MOT (background), $D_{0}$ is the amplitude coefficient of the exponential decay term, and $H(t)$ denotes the Heaviside step function.

The measured scattering rate $R_\text{sc}=4.3_{-2.2}^{+4.1}\times10^6$ s$^{-1}$ is close to the maximum scattering rate for this system, $R_\text{max}=\frac{1}{7}\times1/\tau_{A}=5.9\times10^{6}$~s$^{-1}$ \cite{Metcalf1999}. The value of $R_\text{sc}$ is similar to those measured in atomic MOTs. This observation suggests the possibility of producing strong confining and damping forces, roughly comparable to those in atomic MOTs, if the fraction of scattered photons contributing to the confining force can be greatly increased.

\section*{MOT detection}
The laser induced fluorescence (LIF) collection optics consist of a 150~mm focal-length spherical-singlet lens, followed by a F/0.95 camera lens, then a 650~nm-bandpass interference filter, and finally a CCD camera. The interference filter reflects all repump light at $\lambda_{10}=686.0$~nm, $\lambda_{21}=685.4$~nm, and $\lambda_{32}=684.9$~nm for any angle of incidence (AOI) and transmits $>99$~\% of the $\lambda_{00}=663.3$~nm light at normal incidence; however transmission at $\lambda_{00}$ is reduced for AOI $\gtrsim$ 23$^\circ$.

Using the MOT chamber geometry and assuming the distribution of LIF from the MOT is isotropic, we calculate the geometric collection efficiency of the LIF optics to be $\eta_\text{geo} = 1.1\% $. The amount of light reaching the CCD is further reduced by transmission losses (characterized by $\eta_\text{tra}$) and by AOI-dependent losses of the bandpass filter (characterized by $\eta_\text{fil}$).

We measure $\eta_\text{tra}=0.84$ by tabulating the transmission efficiency of 663.3~nm light through each element of the collection optics at normal incidence. The value of $\eta_\text{fil}$ is measured as follows. Light emission from the MOT is simulated by back-illuminating a thick piece of white Delrin with 663.3~nm light. The front surface of the Delrin is covered except for a 5~mm hole. This creates approximately uniform emission of light over the range of angles incident on the collection optics. The total number of photons hitting the CCD is measured in the presence of all collection optics and again with only the interference filter and CCD present. In this latter configuration, reflection of $663.3$~nm light by the interference filter is negligible since all light is near normal incidence. The ratio of these two numbers is then divided by the ratio of solid angle subtended by the collection optics versus by the CCD sensor alone. Finally, dividing by the transmission losses through the lenses gives the filter transmission efficiency $\eta_\text{fil} = 0.82$ for this geometry. We measure the CCD gain to be $G\approx5.5$~counts/photoelectron and assume the manufacturer-specified quantum efficiency $\eta_\text{qe}=0.53$ for $663.3$~nm light.

The magnification of our imaging system, $M_\text{mag}$, is measured using a grid of black squares back-illuminated with 663~nm light and placed at the appropriate distance from the collection optics. We measure $M_\text{mag}=0.45$, giving a 19.9~mm (horizontal) $\times$ 14.9~mm (vertical) field-of-view.

Due to the high power of 663.3~nm laser light passing through the MOT chamber, scattered light is the primary noise source for the imaging. Several steps are taken to minimize the amount of scattered light reaching the camera. High-quality UV fused silica (10-5 scratch-dig) windows are used on all laser windows. These windows are anti-reflection-coated for 663~nm light and mounted on vacuum nipples far ($\approx\!260$~mm) from the MOT. Scattered light is further reduced by lining the vacuum system with UHV-compatible black copper(II) oxide \cite{Barry2013}. We form and blacken copper sheets in various shapes to line the nipples and the region of the trapping chamber directly in the field-of-view of the camera. Also placed in the nipples are 26-mm-diameter apertures, machined with sharp edges and blackened. At atmospheric pressure, the scattered light is dominated by Rayleigh scattering from air; after pumping down to vacuum, the scattered light signal decreases by $\sim50\times$, to a total detected value of $\approx 1.4\times10^{5}$ photons/ms across the entire field-of-view.

\section*{Trapped molecule number}
The number of molecules observed in the MOT is given by
\begin{equation}
 N_\text{obs} = \frac{N_\text{c}}{G \, \eta_\text{qe} \, \eta_\text{geo} \, \eta_\text{fil} \, \eta_\text{tra} \,  N_\text{per}},
\end{equation}
where $N_\text{c} \approx 7\times10^{5}$ is the (background-subtracted) number of counts registered on the camera over the entire field-of-view for a single pulse of molecules, and $N_\text{per}$ is the number of photons scattered per molecule during the camera exposure. For the default imaging start time and exposure duration, this last factor is given by
\begin{equation}
N_\text{per} = R_\text{sc} \int_{t_\text{im}}^{t_\text{im}+t_\text{exp}} e^{-t/\tau_\text{MOT}} dt \approx 1\times 10^5
\end{equation}
where the integral accounts for the decay of the trapped population (with $\tau_\text{MOT} \approx50$~ms) during the $t_\text{exp}=60$~ms camera exposure. Since MOT loading is essentially complete when the slowing phase ends at $t=40$~ms, and the camera exposure begins $\Delta t=20$~ms later, the initial trapped population is given by
\begin{equation}
N_\text{MOT} = e^{\Delta t/\tau_\text{MOT}} N_\text{obs} \approx 4\times 10^2 \text{ molecules.}
\end{equation}

\section*{Forced MOT oscillation}
The confining and damping forces within the MOT are measured by observing the trapped cloud's response to a rapid displacement of the trap center. Prior to the loading/slowing phase, a shim coil applies a $\approx\!4$~G bias field to offset the trap center by $\Delta\rho\approx \!5$~mm downsream along the axis of the molecular beam. When this bias field is switched off at $t=t_\text{off}$, the center-of-mass of the trapped molecules exhibits damped harmonic motion described by
\begin{equation}
m_\text{SrF}\frac{d^{2}\rho}{dt^{2}}+\alpha \frac{d\rho}{dt}+\kappa_{\rho}\rho = 0.
\end{equation}
Assuming that the cloud is initially at rest, $(d\rho/dt)|_{t=t_\text{off}}=0$, the center-of-mass position vs. time is given by
\begin{equation}
\rho(t) =\rho_{0}e^{-\alpha t/(2m_\text{SrF})}\rm{cos}(\omega_\text{obs}t)
\end{equation}
where $\rho_{0}=\Delta \rho$ is the initial displacement and $\omega_\text{obs}\!=\!\sqrt{\omega_{\rho}^{2} - (\alpha/(2m_\text{SrF}))^{2}}$ is the observed angular oscillation frequency.

\section*{Extracting spatial information using LIF detection}
The weak confinement of the MOT is crucial in order for our LIF-based detection method to extract certain spatial information from the cloud. For the forced MOT oscillation measurement, the camera exposure duration must be short compared to the oscillation period to precisely determine the position of the cloud. We observe $2\pi/\omega_{\rho} = 58(2)$~ms and set $t_\text{exp}=5$~ms; this satisfies the short exposure condition while also allowing the camera to collect enough LIF to accurately measure the spatial distribution.

Similarly, the ballistic expansion measurement uses a camera exposure duration $t_\text{exp}=5$~ms. This duration is short compared to $2\pi/\omega_{z} = 41(1)$~ms, which avoids recapture and compression of the cloud by the trap light during illumination. The maximum free expansion time used, $t_\text{fr}=7$~ms, is capped by the imaging field-of-view rather than the LIF signal-to-noise ratio (in contrast to the case for the MOT oscillation measurement).

\section*{Release and recapture}
An additional measurement of the MOT temperature is performed using a release-and-recapture method. In order to avoid LIF from the untrapped molecular beam, the MOT is released at a fixed release time $t=t_\text{rel}=90$~ms and the free expansion time $t_\text{fr}$ is varied from $0$ to $50$~ms. After each free expansion the MOT is recaptured at $t=t_\text{rel}+t_\text{fr}$, and imaging begins at $t=t_\text{rel}+t_\text{fr}+3$~ms using $t_\text{exp}=30$~ms. In contrast to the free-expansion measurement, this method uses a longer exposure time that gives enhanced sensitivity to the recaptured number of molecules but erases any spatial information about the cloud prior to recapture.

A cloud temperature is determined by comparing the measured recaptured fraction to that of a Monte Carlo simulation, as a function of $t_\text{fr}$. The model assumes isotropic expansion and a spherical trap volume with radius $r_\text{cap}$; molecules inside this radius are assumed to be recaptured with $100$\% efficiency and those outside to be lost. The uncertainty in $r_\text{cap}$ is a well-known limitation of the release-and-recapture method \cite{Lett1988}; we set $r_\text{cap} = d_{\lambda}/2$ to obtain an upper limit on the isotropic temperature. In the Monte Carlo simulation, initial velocities are drawn from a Boltzmann distribution and the effects of gravity are included. Assuming that the MOT is radially symmetric, the initial spatial distribution is inferred from LIF images of the MOT. This procedure gives $T_\text{iso}< 2.7(6)$~mK.

\section*{Diffusion lifetime}
We estimate the lifetime that would be measured for an SrF cloud in the presence of only optical molasses to cross-check the measured values of $\alpha$ and $T_\text{MOT}$ (given the MOT beam diameter $d_{\lambda}$) and to further verify that a trapping force (rather than simply the cooling effect of optical molasses) is necessary to explain our observations. Here the motion of the molecules is treated as Brownian motion within a viscous fluid \cite{Chu1985,Lett1989}. The position diffusion constant $\mathscr{D}_{x}$ is given by
\begin{equation}
\mathscr{D}_{x}=\frac{k_{B}T_\text{MOT}}{\alpha},
\end{equation}
and, using our measured values of $\alpha$ and $T_\text{MOT}$, we calculate $\mathscr{D}_{x}=1.4(3)\times10^{-3}$~m$^{2}/$s. The molasses lifetime $\tau_\text{mol}$ is then given by
\begin{equation}
\tau_\text{mol}= \frac{d_{\lambda}^{2}}{4\pi^{2}\mathscr{D}_{x}}=10(2)\text{ ms}.
\end{equation}
The calculated lifetime $\tau_\text{mol}$ is in agreement with the fits to the data in the presence only of optical molasses. This lifetime is short compared to typical atomic molasses lifetimes (where $\tau_\text{mol}\gtrsim100$~ms) due both to the small damping coefficient $\alpha$ and the relatively high MOT temperature $T_\text{MOT}$. Furthermore, we find the measured MOT lifetime $\tau_\text{MOT}\approx6\times\tau_\text{mol}$, consistent with our observations that molecules are confined in the MOT.

\section*{MOT lifetime}
Although the measured MOT lifetime $\tau_\text{MOT}=56(4)$~ms is short compared to those of typical atomic MOTs, the lifetime is $\sim5\times$ longer than the observed lifetimes of the molasses ($dB_{z}/dz=0$) and damping/anti-restoring ($\mathcal{L}_{00}$ and $\mathcal{L}_{00}^{\dagger}$ polarizations reversed) configurations which have lifetimes of 11(1)~ms and 10(3)~ms respectively (see Fig. \textcolor{blue}{3}c in the main text). It should be noted that these latter two short lifetimes are upper limits due to the temporal and spatial extent of the slowed molecular beam.

As described in the text, it is concluded that the lifetime is limited primarily by ``boil-off" of molecules with energy greater than the MOT trap depth. Before reaching this conclusion, several other possible effects that could limit the MOT lifetime were explored. For example, off-resonant excitation into the ${\rm{A}^{2}\Pi_{1/2}(v'=0,J=3/2)}$ state could lead to decay into the dark X$^2\Sigma_{1/2}(v=0,N=3)$ state. To investigate this loss mechanism, a repump laser was added to the MOT addressing the X$^2\Sigma_{1/2}(v=0,N=3)$ state. The presence of this laser did not change the measured MOT lifetime, indicating that losses due to off-resonant excitations are negligible.

Collisions with He or other background gases have been verified not to be the dominant loss mechanism responsible for the small measured value of $\tau_\text{MOT}$. MOT attenuation may be caused by collisions with residual ballistic He from the buffer gas beam, with background (diffuse) He, or with other gases in the trapping region. We test for attenuation by ballistic He by increasing the flow rate $\mathcal{F}$ of He into the buffer gas beam source from the default value of $\mathcal{F}=5$ sccm to $\mathcal{F}=20$ sccm. This increases the flux of ballistic He incident on the MOT by 4$\times$. In this configuration we measure $\tau_\text{MOT}$ to decrease by only $\sim\!20\%$, suggesting that collisions with ballistic helium are not the primary loss mechanism. With $\mathcal{F}$ still at 20~sccm, we reduce the rotation speed of the turbo-molecular pumps in the trapping region by a factor of $5$, resulting in an increase in all background gas pressures by $\sim5\times$. In this configuration we measure $\tau_\text{MOT}$ to decrease by only $\sim\!25~\%$, indicating that collisions with background gases are not the primary loss mechanism.

\section*{Modelling trap loss}
The measured MOT lifetime $\tau_\text{MOT}=56(4)$~ms corresponds to a total loss rate $1/\tau_\text{MOT}= 18(1)$~s$^{-1}$, $\log(1/\tau_\text{MOT}) = 1.25(3)$. The main loss mechanism is attributed to a shallow trapping potential relative to the MOT temperature, leading to molecules escaping the trap by simply being in the high energy tail of the Boltzmann distribution. Such escape rates depend exponentially on the ratio $U_\text{MOT}/(k_{B}T_\text{MOT})$ \cite{Hanggi1990}. The uncertainties in $U_\text{MOT}$, and to a lesser extent $T_\text{MOT}$, result in predicted loss rates having inherently large associated errors. We have no direct method to measure $U_\text{MOT}$. Instead, $U_\text{MOT}$ is estimated under the assumption that the spring constant $\kappa_\rho$ has a constant value all the way to the edges of the MOT beams. This method is expected to overestimate $U_\text{MOT}$, since the MOT beam intensity is smaller by a factor of $\approx200$ at $\rho = d_{\lambda}/2$ (trap edge) versus at $\rho = 0$ (trap center).

We crudely model the trap loss using a simple Van't Hoff-Arrhenius rate in the low damping limit ($\alpha/(2 m_\text{SrF})\ll\omega_{\rho}$) \cite{Hanggi1990},
\begin{equation}
\frac{1}{\tau_\text{MOT}} = \frac{2\omega_{\rho}}{\pi}e^{-U_\text{MOT}/(k_{B}T_\text{MOT})}.
\end{equation}
Here we multiply the standard 1D prefactor $\omega_{\rho}/(2\pi)$ by a factor of $4$ to account for the two trap edges visited per oscillation and the two radial dimensions; we neglect loss along the deeper axial dimension. Note that the low damping condition is only marginally satisfied, so the prefactor must also be considered as only approximate. Using $U_\text{MOT}/(k_{B}T_\text{MOT})\le4$, this yields an estimated loss rate of $\log(1/\tau_\text{MOT})\ge 0$, an order of magnitude smaller than the measured loss rate. If $U_\text{MOT}$ is instead assumed to have a smaller but realistic value, e.g. consistent with a linear restoring force only out to the measured $1/e^{2}$ radius (7~mm) of the MOT beams, then $U_\text{MOT}/(k_{B}T_\text{MOT})\approx1.6$, and $\log(1/\tau_\text{MOT}) \approx 1.1$ in fair agreement with the measured loss rate. Hence we believe that this model can plausibly account for the loss rate observed in our experiment, although the evidence is not definitive.

\end{document}
%
%
%
%
%
%
%
%
%
%
%
%
%
%
%
%